\documentclass[manuscript]{aastex}
\usepackage{hyperref}

\shorttitle{Transient Sky with VLITE}
\shortauthors{Polisensky et al.}

\begin{document}


\title{Exploring the Transient Radio Sky with VLITE: Early Results}

\author{E. Polisensky\altaffilmark{1}}
\email{Emil.Polisensky@nrl.navy.mil}
\author{W. M. Lane\altaffilmark{1}}
\email{Wendy.Peters@nrl.navy.mil}
\author{S. D. Hyman\altaffilmark{2}}
\email{shyman@sbc.edu}
\author{N. E. Kassim\altaffilmark{1}}
\email{Namir.Kassim@nrl.navy.mil}
\author{S. Giacintucci\altaffilmark{3,4}}
\email{simona.giacintucci.ctr@nrl.navy.mil}
\author{T. E. Clarke\altaffilmark{1}}
\email{tracy.clarke@nrl.navy.mil}
\author{W. D. Cotton\altaffilmark{5}}
\email{bcotton@nrao.edu}
\author{E. Cleland\altaffilmark{6}}
\and
\author{D. A. Frail\altaffilmark{7}}
\email{dfrail@nrao.edu}

\altaffiltext{1}{Naval Research Laboratory, Code 7213, 4555 Overlook Ave SW, Washington, DC 20375, USA}
\altaffiltext{2}{Department of Engineering and Physics, Sweet Briar College, Sweet Briar, VA 24595, USA}
\altaffiltext{3}{Computational Physics, Inc., Springfield, VA 22151, USA}
\altaffiltext{4}{Department of Astronomy, University of Maryland, College Park, MD 20742, USA}
\altaffiltext{5}{National Radio Astronomy Observatory, 520 Edgemont Drive, Charlottesville, VA 22903, USA}
\altaffiltext{6}{Thomas Jefferson High School for Science and Technology, 6560 Braddock Rd., Alexandria, VA 22032, USA}
\altaffiltext{7}{National Radio Astronomy Observatory, P.O. Box O, Socorro, NM 87801, USA}

\begin{abstract}

We report on a search for radio transients at 340 MHz with the Jansky Very Large Array (VLA) Low band Ionospheric and Transient Experiment (VLITE). Between 2015 July 29 and 2015 September 27, operating in commensal mode, VLITE imaged approximately 2800 pointings covering 12,000 deg$^2$ on the sky, sampling timescales ranging from tens of seconds to several hours on a daily basis. In addition, between 2015 February 25 and 2015 May 9, VLITE observed 55 epochs of roughly 2-4 hours each toward the COSMOS field. Using existing radio source catalogs we have searched all the daily VLITE images for transients, while for the COSMOS field we compared individual images and the summed image to search for new sources in repeated observations of the same field. The wide range of timescales makes VLITE sensitive to both coherent and incoherent transient source classes. No new transients are found, allowing us to set stringent upper limits on transients at milliJansky levels and at low frequencies where comparatively few such surveys have been carried out to date. An all-sky isotropic rate of bursting radio transients with similar rates, duration, and intensity as the unusual transient GCRT J1745$-$3009, discovered in wide-field monitoring toward the Galactic center, is ruled out with high confidence. The resulting non-detections allows us to argue that this is a coherent source, whose properties most resemble the growing class of nulling pulsars. We end with a discussion of the future prospects for the detection of transients by VLITE and other experiments.

\end{abstract}


\section{Introduction}

In recent years there has been an explosive increase in the number of radio telescopes operating at MHz radio frequencies, enabled by technical advances in aperture synthesis arrays, digital electronics, signal processing, and high speed networking computing \citep{g13}. The systematic exploration of the dynamic radio sky is a core science goal for most of these facilities \citep{fb11}. While interesting new discoveries have been made in the time-domain at MHz frequencies \citep{oth+14, sfb+16}, large numbers of new transients have not materialized \citep{bmk+14, cvw+14}.

The transient and variable radio sky at MHz frequencies is different than the variable GHz radio sky, and optimal strategies for observing one may not apply to the other. For example, propagation effects through the interstellar (and interplanetary) medium become more pronounced at longer wavelengths \citep[e.g.][]{ktm+15,phs+15}. The flux density variability of the AGN (that dominate the quiescent sky) become driven by refractive interstellar scintillation at these frequencies and not by the intrinsic variations seen at higher frequencies \citep{mdc+94}.   The types of sources that dominate the radio sky at MHz frequencies are also different from those in the GHz sky. Some types of sources become easier to observe at lower frequency (e.g. pulsars, flare stars), while others are more difficult to detect (e.g. supernovae). 

Both general considerations \citep{mwb15} and empirical evidence \citep{pfk15} show that for incoherent synchrotron emission the brightness temperature is limited, and thus the more luminous transients evolve on long (days/years) characteristic timescales.  These luminous transients are in principle detectable to large distances (and hence large volumetric rates), but synchrotron self absorption and thermal absorption both conspire at MHz frequencies to push the required cadence to years not days.  Coherent emission, on the other hand, has no restrictions on the maximum brightness temperature and can thus be detected on much shorter timescales.  For detecting known coherent sources classes (fast radio bursts, pulsars, etc.) the experimental parameters can be matched to the known source duration and brightness \citep{sha+11, srb+15, ttw+15}.   Searching for new source classes is more difficult, because the range of possible luminosities and timescales is large \citep{c08}, and the physics of coherent emission is not understood well enough to constrain it. For a fixed amount of observing time, wide-shallow surveys will yield larger detection rates than narrow-deep surveys \citep[e.g.][]{mhb+16}, but the integration timescales over which to search for the coherent emission are largely unknown.


Recognizing these constraints at MHz frequencies, we have designed a large etendue\footnote{The etendue of a telescope is defined as the product of its collecting area (A) times the field of view ($\Omega$).} experiment that is sensitive to both incoherent and coherent transients on a range of timescales. The Jansky Very Large Array (VLA) Low band Ionospheric and Transient Experiment (VLITE) is a commensal system that operates continuously at the prime focus of the 25-m antennas using the low frequency receivers while the main science program is being conducted with the Cassegrain optics at higher frequencies. Operating in this mode, a single pointing of VLITE images 5.5 $\deg^2$ enabling tens of thousands of square degrees to be imaged over the course of a month. Integration times vary from tens of seconds to several hours while repeated visits to the same field give cadences of day-long to year-long timescales. Using existing catalogs and archival data we can increase the cadence to decade-long timescales.  

This paper is organized as follows. In \S\ref{system} we provide a short overview of the VLITE commensal system, with an emphasis on aspects relevant to radio transients. In \S{\ref{obs} we describe our daily integrations and the deep pointings toward the COSMOS field. We describe our results and their interpretation in \S\ref{results}, and we summarize the experiment with a look towards its potential evolution in \S\ref{conclusions}.

\section{The VLITE System}\label{system}

In late 2014 the Remote Sensing Division at the US Naval Research Laboratory (NRL) secured funding for a commensal experiment that would operate using the newly re-designed low frequency system at the VLA \citep{cla11}. The scientific motivations included astrophysical imaging and serendipitous transient radio astronomy as well as ionospheric remote sensing.  VLITE was conceived as a prototype 10-antenna system with dedicated samplers and fibers that tap the signal from 10 of the new VLA P-band receivers and correlate them through a dedicated DiFX-based software correlator \citep{deller07}. A 64 MHz bandwidth centered on 352 MHz is divided into 100 kHz frequency channels with full polarization taken at 2 second sampling. The useable bandwidth is restricted to an RFI-free 40 MHz centered at 340 MHz due to interference from the US Navy Mobile User Objective System (MUOS) satellites that use the UHF band (300-400 MHz) for communications.   Full technical descriptions of the VLITE system for radio astronomy and ionospheric remote sensing are given in \citet{cla16,cla16b} and \citet{hel16}. We focus here on those parts specific to transient detection. 

VLITE began full science operations in November 2014 and it is expected to be operated across two full cycles of the extended A and B configurations. VLITE operates nearly continuously and independent of the VLA on-line system except that its pointings are slaved to the Cassegrain science program. In a one hour integration in B configuration the rms sensitivity is typically $\sim 3.5$ mJy beam$^{-1}$. The instantaneous field of view (to the primary beam full width half power) at this frequency is $\sim 5.5$ $\deg^2$ and the data rate is 2 GB/hour.  All of the data are archived at NRL. 

At the end of each UTC calendar day the VLITE data are copied over from disk at the VLA site via the network to NRL. The data are then bulk processed by the VLITE Astrophysics pipeline processing, which combines standard tasks from both {\it AIPS} \citep{gre03} and {\it Obit} \citep{cot08} data reduction software. 

Due to the physical configuration of the VLA antennas, the low frequency receivers move to a slightly different position for each of the Cassegrain receivers.  As a result, the gain and bandpass shape of the VLITE data vary with the frequency, or primary observing band, in use by the telescope's main science program.  The data for each band must be calibrated separately.   IDIFits scans for each UTC day are loaded into {\it AIPS} and sorted into datasets based on the primary observing band and the antennas in operation.  All further processing proceeds on these ``primary band" datasets.  

The data are flagged using automated tasks, and all frequencies above 360 MHz are removed completely due to the persistent MUOS emission. All observations of a set of 7 primary calibrators (3C\,48, 3C\,138, 3C\,147, 3C\,196, 3C\,286, 3C\,295, 3C\,380) are used to solve for the delay, gain, and complex bandpass, using models distributed with {\it Obit}.  After calibration, the data are flagged a second time, and then re-calibrated. The {\it AIPS} task `FACES' is used to generate a sky model for each pointing from the NRAO VLA Sky Survey catalog \citep[NVSS:][]{ccg+98}. The data are phase-calibrated to these models using a solution interval of 12 seconds, and the solutions are smoothed to one minute and applied to the data; solutions which fail are replaced by interpolated values from neighboring time samples preventing data loss during this step.  

Images are made using the {\it Obit} task `MFImage', which divides the data into three roughly equal frequency sub-bands, and is set to perform two imaging and phase self-calibration cycles before a final image is created.  The task uses an ``autobox'' setting which identifies and boxes emission in the image for cleaning.  The full 2.5$^\circ$ FWHM of the primary beam is covered with facets, and outlier facets are placed on bright sources to a radius of 20$^\circ$.  The result is an image cube with three frequency planes as well as a combined ``average" final image which can be used for transient searches.   

Because we combine data from an entire day of observations, image sensitivity for VLITE is dependent not just on integration length (time on source), but also on the total time spanned by the observations used in the image.  The sparse, 10-antenna VLITE array provides limited UV-coverage and poor snapshot sensitivity.  Images made from many short integrations spread out over the course of the day benefit from the improved UV-coverage.  To make the distinction clear, we always refer to the time on source as the {\it integration} time, and the total time spanned by the observations as the {\it duration}.  

\section{Observations}\label{obs}

We identified two subsets of the VLITE data to use for this preliminary investigation of transients.  The first were a set of all images made during a period of two months in late summer of 2015.  These daily images can be compared to existing radio catalogs to identify any new sources not previously known.  We discuss these data in section \S\ref{daily}.   The second data set are taken from two programs which combined to spend hundreds of hours making deep observations of the COSMOS field.  Because the observations repeat we can compare them to each other to search for variability and transients on timescales of days and months.  These data are discussed in section \S\ref{cos}.

\subsection{Daily Images}\label{daily}

During the summer of 2015 the VLA was in A configuration with maximum baselines of 36 km.  The combination of long baselines with the sparse VLITE array made reliable automated imaging over the entire primary beam to FWHP difficult.  A decision was made in July 2015 to limit VLITE to maximum baselines $\sim 15$ km, in a configuration which we have designated ``B+'' because the baselines are all slightly greater than those in a true VLA B configuration.  Between 2015 July 29 and 2015 September 27 VLITE observed, on average, 50 independent directions per day, for a total of 3713 pointings over this period. For the purposes of this experiment, observations of the same direction made at different primary bands or with different antenna sets are treated as independent pointings.  The images were examined daily and compared to the NVSS catalog visually to catch potential transients in near real-time.

In the autumn of 2015 we developed software to automate comparison of the {\it Obit} catalog of sources exceeding 10 times the local rms for each image to the Westerbork Northern Sky Survey \citep[WENSS:][]{rtb+97} and NVSS catalogs. If a VLITE source is within one arcminute of a catalog source it is classified as a matched source, if not, it is classified as unmatched and a potential transient candidate. We searched only for previously uncataloged sources and not for variable sources or previously cataloged long-timescale transients that have since faded. Of the 3713 images for this period, 78 are centered at declinations $\delta < -40^\circ$, which is below the coverage limit of the reference catalogs. Another 807 images are failed images containing no sources above $10\sigma$ significance with 29 additional images visually identified as unsuitable for searching due to large numbers of imaging artifacts. The remaining 2799 images constitute our good image set. The total integration time of these images is $\sim 575$ hours.

In Fig.\,\ref{fig1} we show the distribution of our set of 2799 pointings on the sky, color-coded to distinguish the different integration times.  Many of the deeper integrations are calibrator fields that are observed multiple times each day but some also include targeted deep observations.  

The distribution of image duration times for our good image set is shown in the left panel of Fig.\,\ref{fig2}.  For VLITE transient science, the duration is a more important timescale than the integration because it represents the total timescale in the image.  Typical duration times for the daily images ranged from 30 to 2000 seconds. The drop in the number of images at short duration highlights the difficulties in snapshot imaging with a sparse array like VLITE.  Nearly one in five of the 3713 daily images has a duration $< 100$ sec but nearly four out of five of these are rejected as bad images with no significant sources. 

The distribution of rms noise is shown in the right panel of Fig.\,\ref{fig2}. The typical rms noise values ranged from 3 to 50 mJy beam$^{-1}$. As can be seen in Fig.\,\ref{fig3}, the general trend is for the rms noise to decrease with the square root of the duration time, however, even after discarding obviously bad datasets there are images with large excursions in the rms noise regardless of the duration time. While we expect some amount of scatter due to different amounts of integration time for images of the same duration, particularly for the longer duration images, imperfect subtraction of RFI in the pipeline processing also increases the rms noise.

There were $52,985$ sources in our set of 2799 images. The majority of these were matched to $13,735$ unique catalog sources, however, 924 sources in 444 images were unmatched without WENSS or NVSS counterparts. Seven of these unmatched sources were in low-noise images and matched to known weak radio sources with the NASA/IPAC Extragalactic Database (NED). Six sources were detections of two known pulsars, B1859+03 and B1919+21. Visual investigation of the remaining unmatched sources in the multi-frequency images revealed the majority to be imaging artifacts, however, 32 could not definitively be identified as artifacts and were promoted to candidate transient sources. Automated and manual reprocessing of the candidates revealed all 32 to be imaging artifacts. We highlight the two most common candidate types below.

Figure \ref{Candidate1} shows a typical candidate transient source identified in the original pipeline image on the left and a larger field of view followup image on the right.  This observation was pointed at IRAS 16552-30 and the candidate transient is $\sim 1.25^\circ$ from the pointing center.  The field was observed for nearly 10 minutes on 2015 August 19 as part of program VLA/15A-301.  The original image has a resolution of $28.5\arcsec$ (the original image was made at lower resolution for the sake of processing speed), while the remade image is at a resolution of $12.3\arcsec$. The two images have comparable rms values $\sigma \sim 11$ mJy bm$^{-1}$ in this area.  Sources which have an NVSS match are circled small in cyan.

The 11$\sigma$ candidate, circled large in green, lies quite close to the edge of the original image and is likely the result of the automated imaging attempting to CLEAN sidelobes from a real source which lies just outside of the original image area.  The real source does have a match in the NVSS catalog but it was not bright enough to have an outlier image facet placed onto it.  Nearly a third of our candidate transients were found at the edge of the imaging area and disappeared when a larger field-of-view image was made with automated imaging.  

This particular image also illustrates why we do not look for transients below 10$\sigma$ \citep{fko+12}. There are a number of 7-8$\sigma$ ``sources" clearly visible which do not match NVSS sources (do not have small cyan circles around them), in addition to one or two which do have NVSS counterparts.  The fake sources are mostly a result of automated CLEAN trying to put sources on sidelobes of VLITE's messy dirty beam; in this particular case the problem was exacerbated by the lower resolution imaging process and was significantly less of a problem in the remade image.  The number of false positives increases substantially when we try to search for transients in the daily images at lower significance than 10$\sigma$.

Artifacts in the proximity of bright sources account for another third of the transient candidates. While not uncommon to the VLA generally, the situation is exacerbated for VLITE due to high sidelobes associated with a dirty beam derived from $\leq$10 antennas. An example of this phenomena is provided in Fig.\,\ref{Candidate2}. This observation was pointed at J1851+0035 and the candidate transient is next to the target source at the pointing center.  The field was observed for $\sim4$ minutes on 2015 July 31 as part of program VLA/15A-301.  Both the pipeline and the reprocessed image have a resolution of $12.3\arcsec$, and a local rms of 21 mJy bm$^{-1}$. The left panel is a cutout from the pipeline processed image centered on the bright source J1851+0035. The transient candidate to its southwest is a $>14 \sigma$ detection and circled in green. The central image shows the same region after manual reprocessing. The right panel shows the dirty beam sidelobe pattern for this field.  It reveals that the candidate corresponds to one of the peaks in the sidelobe pattern. The automated imaging procedure in the pipeline incorrectly CLEANed this sidelobe; when it is excluded from the CLEAN area in the manual processing the candidate no longer appears.  Limiting the pipeline CLEAN to small areas around known sources in future processing should greatly reduce the number of  false candidates from sidelobes.

\subsection{The COSMOS field}\label{cos}

The Cosmic Evolution Survey (COSMOS) covers a 2 deg$^{2}$ region centered at RA=10$^{h}$, Dec = +02$^{\circ}$ \citep{sab07}.  This region has been chosen as the focus of a deep multi-wavelength study to probe the evolution of extragalactic source populations, dark matter, and large scale structure over the redshift range $0.5<$z$<6$.  The survey provides large samples of high redshift objects, and includes spectroscopy and imaging at wavelengths from the X-ray to the radio regimes.   

During the B configuration (maximum baselines of $\sim11$ km), the VLA observed the COSMOS field for several hundreds of hours of integration time under two observing programs.  Between 2015 February 12 and 2015 May 9, a project to study cosmic star formation at sub-$\mu$Jy sensitivity made 100 hours of observations at S-band \citep{cosmos}.  Between 2015 February 25 and 2015 May 4, the COSMOS HI Large Extragalactic Survey (CHILES), a project to probe HI emission out to redshifts of z$\sim 0.5$ made 252 hours of observations at L-band \citep{fgh+13}\footnote[1]{a description of CHILES is available at \url{http://chiles.astro.columbia.edu}}. 

The pointing centers for the two projects were offset from each other by approximately $20\arcmin$, but there is a substantial overlap area within the large VLITE primary beam, which also completely covers the COSMOS field area. Both of these surveys repeatedly observed the same position on the sky for several hours integration each day spaced over several months. These fields present an opportunity for VLITE to search for fainter transients and on shorter timescales than is possible with the daily observations where we must rely on comparison with existing surveys that push the timescales to years. 

Observations for these data were reduced using the pipeline software as described in Section \S\ref{system}, with the exception that they were not combined with any other data taken during the same day. This was motivated by the fact that the CHILES project has a unique correlation mode in VLITE with double the spectral resolution and it cannot be combined with any other projects during primary calibration.  We chose to isolate the S-Band data from the rest of the daily observations as well.   The main effect is that each was calibrated using only its own primary calibrator, 3C\,286, leading to a more uniform flux scale.  Slightly larger {image}s (3$^\circ$ across) were made to allow reliable source searching over the entire primary beam FWHP field of view.  A few of the VLITE datasets failed to calibrate, but we obtained 34 images of the CHILES pointing and 21 images of the S-Band pointing, with a combined total of 225 hours of integration.  The average integration length for each image is $\sim 4$ hours.

The images for the CHILES and S-Band fields were combined in the image plane to create a single ``best" image for each pointing.  A global rms was measured from a fit to the histogram of the pixels in the inner quarter of each image. Images with global rms values greater than 2.5 mJy/bm were removed from the combination and the remaining maps were weighted by the inverse of their global rms values. The final CHILES image includes 27 individual 4-hour images while the S-Band image includes 11; both have a global rms that is just under 1 mJy beam$^{-1}$.  The noise in these combined maps is dominated by artifacts around bright sources and sidelobes from unCLEANed faint sources.  In order to identify candidate transients we ran the LOFAR transient pipeline \citep[TraP:][]{ssm+15}, first on the combined image to set up a reference catalog and then on all of the daily images, searching out to maximum radii of $1.34^\circ$ and $1.20^\circ$ for CHILES and S-Band, respectively.

Unlike our search of the daily images, which relies on previously published survey catalogs, the TraP software builds its own catalog of sources and uses metrics of the light curves to flag transient candidates. Extensive testing of the TraP parameters established that a detection threshold of $10\sigma$ provides a good balance between detecting real sources and limiting the number of candidate transients to a manageable number for manual followup.  Because of the offset pointing centers, the CHILES and S-Band fields were processed separately, resulting in about 130 sources in the TraP running catalog for each field with 20 transient candidates across both fields. Examination of the VLITE multi-frequency images revealed all candidates to be sporadic artifacts around bright sources that achieved $> 10\sigma$ on a single image.

\section{Results and Discussion\label{results}}

Our preliminary VLITE sample for transient searches includes 2854, 5.5 deg$^2$ images at random directions on the sky above $\delta >-40^\circ$, which combined include $\sim$800 hrs of total on-sky integration time spread over 4.5 months of observations. We have searched for new radio sources which appeared in our images but were not previously known in existing radio source catalogs and for new sources that appeared in repeated pointings of the same fields. No image-plane transients were detected at a typical 10$\sigma$ flux density threshold of 100 mJy beam$^{-1}$.

Timescale is important in estimating the sensitivity of VLITE to transients. The observational phase space for physically-interesting MHz transients is large. As noted by \citet{cvw+14} and many others, a transient survey over some sky area with a given flux density sensitivity limit must also be characterized by the duration of each pointing and the cadence at which the observations are made. Whether or not a given source class can be detected depends on these timescales. This preliminary VLITE experiment does not probe variability on the sub-second timescales of pulsars, fast radio bursts (FRBs) or rotating radio transients (RRATs), but instead is sensitive to sources such as magnetars, flare stars and long gamma-ray bursts (LGRBs), which vary on timescales of minutes, hours, days and years.

To compare our results with transient event rates and surface densities from other studies we assume the transient source populations are isotropic and have a surface density that traces a Euclidean distribution.  We calculate upper limits for two typical timescales present in our data; a transient that is visible for 10 minutes and another that is visible for 6 hours. For each of these two timescales we estimate the number of epochs at each flux density that VLITE would have detected a signal ($\geq 10\sigma$) out to the half power of the VLA primary beam. For epochs with duration times longer than the characteristic transient timescale we have reduced the signal-to-noise to allow for time-averaging of the peak flux.  However, the search of a single image made from an epoch sensitive enough to detect a time-averaged signal is an implicit search for non-averaged signals over each sub-epoch, assuming that the noise does not vary significantly over time. Taking each sub-epoch into account would decrease the VLITE surface density upper limit by, for example, a factor of $\sim 2$  at 1 Jy for transients with a 10 minute timescale.  We have not included this decrease, however, since we did not actually divide each VLITE observation into sub-epochs.

The limits we calculate are tabulated in Table \,\ref{tbl-1} and plotted as orange (10 minute) and red (6 hour) curves on the surface density plot in Fig.\,\ref{fig6}.  A series of diagonal lines show estimated rates for several of the more common known classes of MHz radio transients and circles with error bars are plotted for rates from previous detections of transients of unknown type \citep{2012AJ....143...96J,hlk+05,sfb+16}.  In addition, we mark the upper limits from several published transient searches with no detections.

The event rates for the known Galactic and extra-Galactic transient source populations are taken from the literature \citep{wbc+13,mwb15,mhb+16} and scaled by $\nu^{-0.7}$ (i.e. an optical thin synchrotron spectrum) where necessary. The uncertainties on these rate estimates are generally larger than any frequency adjustments. Not all possible known source classes are plotted on Fig.\,\ref{fig6}, just those that are expected to significantly contribute to the all-sky rate at MHz frequencies on the timescales that the VLITE experiment is sensitive. Flares from dwarf M stars likely dominate the Galactic all-sky rate \citep{wbc+13}. At MHz frequencies the brightness temperature limit set by inverse Compton cooling makes incoherent emission mechanisms (cyclotron, gyro-synchrotron, synchrotron) unlikely. \citet{mwb15} show that the extragalactic populations that dominate the MHz transient sky are likely magnetar-producing neutron star mergers, off-axis tidal disruption events, and, to a lesser extent, black hole-producing neutron star mergers and long gamma-ray bursts. At 350 MHz the evolutionary timescales for all of these transients are large (200-2000 days) because they are the result of synchrotron emission. 

Surprisingly, the three rates calculated from detections of transients from {\it unknown} sources in Fig.\,\ref{fig6} are all much higher than the known radio source transient rates in the MHz sky. As a result, the upper limit to the all-sky rate of MHz transients is set, over a wide flux density range, by unknown source types.

The first of the three unknown-type transient detection rates is taken from \citet{sfb+16}, based on repeated observations of the North Celestial Pole with LOFAR \citep{lofar}.  Their data includes 2149, 11-minute snapshot images, each covering 175 deg$^2$. The transient was detected in a single 11-min image as a $\sim$15-25 Jy event at 60 MHz.  The two months of VLITE images included in this preliminary study do not cover enough of the sky to say anything interesting about the 340 MHz properties of the 15-25 Jy LOFAR transient seen at 60 MHz.

We also plot a result from \citet{2012AJ....143...96J} who searched a 6.5 deg$^2$ region with six 12-hr VLA observations taken at 325 MHz toward the Spitzer Space Telescope Wide-Area Infrared Extragalactic Survey \citep[SWIRE:][]{swire} Deep Fields with cadences from 1 day to 3 months. In one 12 hour epoch they found a single transient that was weak in the first 6 hours but reached a peak flux density of 2.1 mJy during the subsequent 6 hours. Our current VLITE images are not sensitive to flux densities of this magnitude on 6 hour timescales and our limits for detection of 6 hour transients, indicated by the red curve on Fig.\,\ref{fig6}, lie above the Euclidean extrapolation of the maximum allowed rate from this detection.  

Finally we plot the rate calculated from the detection of four radio transients with the VLA at 330 MHz and the Giant Metrewave Radio Telescope \citep[GMRT:][]{gmrt} at 330 MHz and 235 MHz while monitoring the Galactic center region. Three of these are known as the Galactic center radio transients (GCRT). These include GCRT J1746$-$757 \citep{hlk+02}, the bursting transient GCRT J1745$-$3009 \citep{hlk+05}, and GCRT J1742$-$3001 \citep{hwl+09}. GCRT J1746$-$2757 is a 200 mJy source detected in only one observation while the other two GCRTs were detected multiple times with flux densities in the range of 0.1 to 1 Jy.  In addition to these three radio-discovered events, the radio counterpart to the X-ray transient XTE\,J1748$-$288 was seen in the same monitoring program.  Excluding repeated detections of the same transient, \citet{mhb+16} report a rate for these sources of 3.6$\times 10^{-3}$ deg$^{-2}$ at 50 mJy based on continued monitoring of the Galactic Center through 2015. 

One of these four sources, GCRT J1745$-$3009 (the ``Burper'') is a repeating transient radio source that has been the focus of many studies. A series of five $\sim 1$ Jy bursts, each with a duration of $\sim 10$ min, and occurring at intervals of $\sim 77$ min were initially seen in 2002 \citep{hlk+05}. Single bursts were also detected in 2003 and 2004 with GMRT observations \citep{hlr+06,hrp+07}. In both GMRT detections, however, the observation is too poorly sampled to know if the 77 min burst recurrence interval was present. At least one of the GCRT J1745$-$3009 bursts exhibits time-varying, highly circularly polarized emission that approaches 100\% \citep[][]{2010ApJ...712L...5R}. Several of the bursts appear to have a steep spectrum, in one case with $\alpha \sim -13$ (where S$_\nu\propto\nu^{\alpha}$), derived across the 30 MHz GMRT bandpass \citep{hrp+07,spr09}. 

The distance to GCRT J1745$-$3009 is not well constrained due to a lack of an associated counterpart at either optical/NIR wavelengths or at high energies. If it is located at or near the Galactic center, with d$>>$70 pc, the brightness temperature limit for it exceeds the 10$^{12}$ K limit set by inverse Compton cooling for incoherent sources.   In that case the emission must be coherent in nature and GCRT J1745$-$3009 must be a rare event at or near the center of the Galaxy.   However, if GCRT J1745$-$3009 is nearby (d$<$70 pc), it is most likely a late-type magnetically-active star which flares periodically \citep{2002ARA&A..40..217G}. Such sources are expected to be distributed throughout the sky.  As noted by \citet{kp05}, wide-field surveys, such as VLITE, should expect to see many sources similar to GCRT J1745$-$3009 in that case.  

We estimate the event detection rate and surface density for GCRT J1745$-$3009 which can be compared to the VLITE upper limits.  In $208$ hrs of monitoring and archival observations ($\sim 30$ min to $\sim 7$ hr duration) toward the Galactic center at 330 MHz with the VLA and 330 and 235 MHz with the GMRT, seven events were detected from GCRT J1745$-$3009; five 1 Jy ten-minute long bursts in 2002, a single 0.5 Jy burst in 2003, and a weak 50 mJy signal in 2004. We exclude the 50 mJy signal because it is too weak for detection by VLITE. Our sample is thus six bursts of $\sim 10$ minute duration and $\sim 1$~Jy peak flux density, detectable within a FOV out to a radius of one quarter of the primary beam response of each instrument, for each observing frequency. Accounting for the observing time and FOV of each dataset we calculate an event rate of $3.6\times 10^{-3}$ hr$^{-1}$ deg$^{-2}$. The data were also imaged and searched on $10$~min intervals for which we derive a surface density estimate of $6.0\times 10^{-4}$ deg$^{-2}$ above 0.8 Jy.  This result is plotted on Fig.\,\ref{fig6}.

The VLITE upper limits for transients with a 10 minute duration similar to the Burper are indicated by the orange curve from 0.1 to 100 Jy on Fig.\,\ref{fig6}.  The GCRT J1745$-$3009 95\% lower limit lies above the VLITE 95\% upper limit at $\sim 1$~Jy. In this experiment, about 2200 VLITE observations with a total integration time of $402$ hours had sensitivity to Jy-level bursts on 10 minute timescales over an area of 5.5 deg$^2$.  If the Burper represents an isotropically distributed source population, we would expect to see at least 8 detections in the VLITE data. None were found. The probability of such an occurrence is only $0.03\%$. We  thus conclude an isotropic population with similar rates, duration and intensity as GCRT J1745$-$3009 does not exist. Any such population must be at least an order of magnitude less common, with the current limits set by the 11-min LOFAR transient or the 6-hr $\sim$mJy VLA transient.

Relaxing the isotropic assumption disfavors local M-star or dwarf models \citep[e.g.][]{2010ApJ...712L...5R} for the Burper, and pushes its distance beyond 100 pc. This result, together with the non-detection of fainter bursting transients toward the Galactic center, increases the likelihood that GCRT J1745$-$3009 lies close to the Galactic Center and requires that the emission be coherent in origin \citep{hlk+05}. Pulsars are a coherent emitter that can satisfy all the observational properties of GCRT J1745$-$3009. More exotic models have been suggested such as a precessing pulsar and a white dwarf ``pulsar" \citep{zg05,zx06}, but the simplest hypothesis is that GCRT J1745$-$3009 is a nulling pulsar \citep{kp05}. 


\section{Conclusions}\label{conclusions}

We have used VLITE, the new 340 MHz commensal observing system on the VLA, to search for astrophysical transients on timescales from tens of seconds to several hours. Our analysis included daily images obtained over a period of two months with a total integration time of 575 hours and covering a wide range of durations, and 225 hours of deep, $\sim$4 hr duration observations on a single pointing. We detected no transients. 

Coupled with earlier results, our current VLITE limits show that the surface density of detected, unidentified MHz transients dominates predictions derived from all known populations. These unknown transients set the all-sky rate for MHz transients and include sources ranging in flux density from a few mJy to over 10 Jy. At the mJy level, the surface density remains set by \citet{2012AJ....143...96J} for transients of $\ge$~6 hr duration. The current analysis cannot improve constraints on transient rates set by that and by the $\sim$11 minute duration, bright (15-25 Jy) transient reported by \citet{sfb+16}.  

For intermediate flux densities and durations, the current surface density of MHz transients is dominated by the GCRTs, the brightest of which is the repeating source GCRT J1745$-$3009. Comparison to limits from our current VLITE experiment show that it cannot be part of an isotropic source population.  We conclude that GCRT 1745$-$3009 is most likely to be located close to the Galactic Center. The corollary, based on its brightness temperature, is that it must be coherent, as originally suggested by \citet{hlk+05}. The most likely scenario remains its association with a compact rotating object such as a nulling pulsar.

Ongoing MHz transient searches include efforts by dipole-based instruments at the MWA \citep{bmk+14} and LOFAR \citep{sfb+16}, as well as by dish-based telescopes like the GMRT \citep{hwl+09} and VLA \citep{hlk+02}. The dipole based arrays have naturally large fields of view and lots of observing time, though their sensitivity is often confusion-limited and their newer systems are generally less well characterized. The dish-based systems are older and better understood, but have smaller fields of view and much less dedicated observing time. VLITE occupies a unique niche, combining 1) a moderate size field of view, especially for a dish based telescope, 2) good sensitivity on a well characterized instrument, and 3) copious observing time. Moreover its observing cadences are relatively unbiased with respect to the duration of potential transients.

Our experience with the data analyzed so far allows us to make predictions on two levels. Firstly, we project the results from 1 year of VLITE observations based on the data analyzed in this paper. We additionally project the results of breaking {\it all} longer duration observations into shorter ones, e.g. existing and future 4 hour COSMOS observations will also be analyzed as 24 distinct 10 minute durations. Fig.\,\ref{fig7} indicates the resulting 1 year limits, assuming no VLITE detections. Ignoring various subtleties, to first order the VLITE 1 year curves drop downward, especially the 10 minute curve that drops below the 6 hour curve at thresholds above $\sim$200 mJy. This significant improvement indicates the underlying importance of analyzing all future VLITE data on finer timescales. Additional innovations in our data reduction can only make our limits stronger. For example, we may expand the field of view cataloged for each pointing to increase the area of the experiment. Note that doubling the field of view would halve the 10 minute surface density limits at the highest thresholds, coming within a factor of $\sim3$ of the \citet{sfb+16} detection. This improvement is feasible, especially in the more compact configurations of the VLA.

VLITE development was undertaken with a vision of transitioning the narrow-band ($\Delta \nu =$ 64 MHz), 10 antenna system into a full VLA capability across all 27 antennas accessing the full bandwidth of the P-Band system ($\Delta \nu =$ 256 MHz). This proposed LOw Band Observatory, LOBO, would be similar to VLITE and could run on all traditional observing programs.  However, it would provide significantly increased imaging fidelity, much lower rms noise levels, and broader spectral coverage. 

The expansion of VLITE to LOBO will have a major impact on MHz transient detection because it will be possible to detect known classes of MHz transients. This is indicated on Fig.\,\ref{fig7} that also includes LOBO one year limits. On 10 minute timescales, LOBO is expected to detect dMe flare stars. For longer ($\geq$6 hours) durations, LOBO should begin to detect weaker sources in the population defined by the \citet{2012AJ....143...96J} detection, and on even longer timescales, LOBO will be sensitive to neutron star mergers that create magnetars.  Even off-axis TDEs may be within its reach. The projected LOBO limits would also rule out an isotropic distribution for the fainter ($<$200 mJy) non-bursting GCRTs, as well as for fainter and/or shorter duration bursting transients, e.g. like the one 50~mJy, 2 min burst observed from GCRT J1745$-$3009. 

We also include projected limits from 2 minute durations for one year of LOBO observations. While some short ($\leq 2$ minute) VLITE images were used in our analysis, many more did not contain sources with $> 10\sigma$ significance due to sparse snapshot UV-coverage. This situation will improve dramatically for 27 antenna LOBO snapshots, augmented by richer bandwidth synthesis and better sensitivity. Thus the domain of short duration transients will be much more efficiently sampled by LOBO. 


Finally we note that the VLITE design also incorporates a GPU-based direct voltage capture system for fast (msec) transient detection, including pulsar giant pulses, RRATs, and possibly even FRBs. While many fast transients will not be detectable at meter-wavelengths due to propagation effects (e.g. scattering and absorption), the advantage of arc-second localization nonetheless render the capability attractive for steep-spectrum and coherent phenomena, especially for commensally obtained data.




\acknowledgments

The National Radio Astronomy Observatory is a facility of the National Science Foundation operated under cooperative agreement by Associated Universities, Inc. Basic research in radio astronomy at the Naval Research Laboratory (NRL) is funded by 6.1 Base funding. This work made use of data from the VLA Low band Ionospheric and Transient Experiment (VLITE). Construction and installation of VLITE was supported by the NRL Sustainment Restoration and Maintenance funding. SDH thanks Dr.\ Huib Intema at NRAO for low-band data reduction. DAF thanks Dr.\ Chryssa Kouveliotou at GWU for her support during the time in which this paper was begun. This research has made use of the NASA/IPAC Extragalactic Database (NED) which is operated by the Jet Propulsion Laboratory, California Institute of Technology, under contract with the National Aeronautics and Space Administration. 

{\it Facilities:} \facility{VLA}.




\clearpage

\begin{figure}
\epsscale{0.80}
\plotone{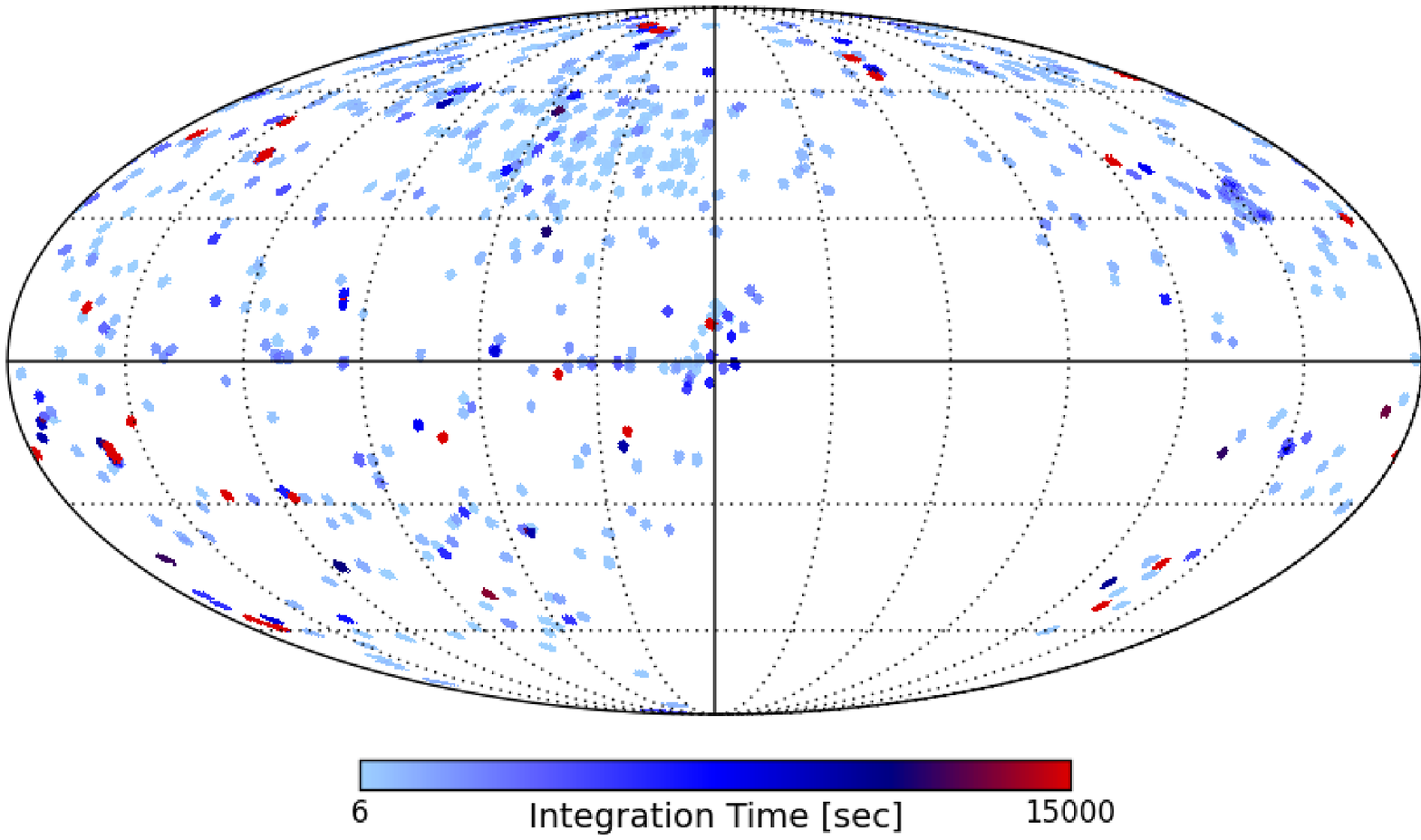}
\caption{Aithoff projection showing the distribution of the 2799 daily VLA pointings over the two month interval and 55 pointings of the COSMOS field. Fields are colored by their total integration time across all images.}
\label{fig1}
\end{figure}
\clearpage

\begin{figure}
\epsscale{1}
\plottwo{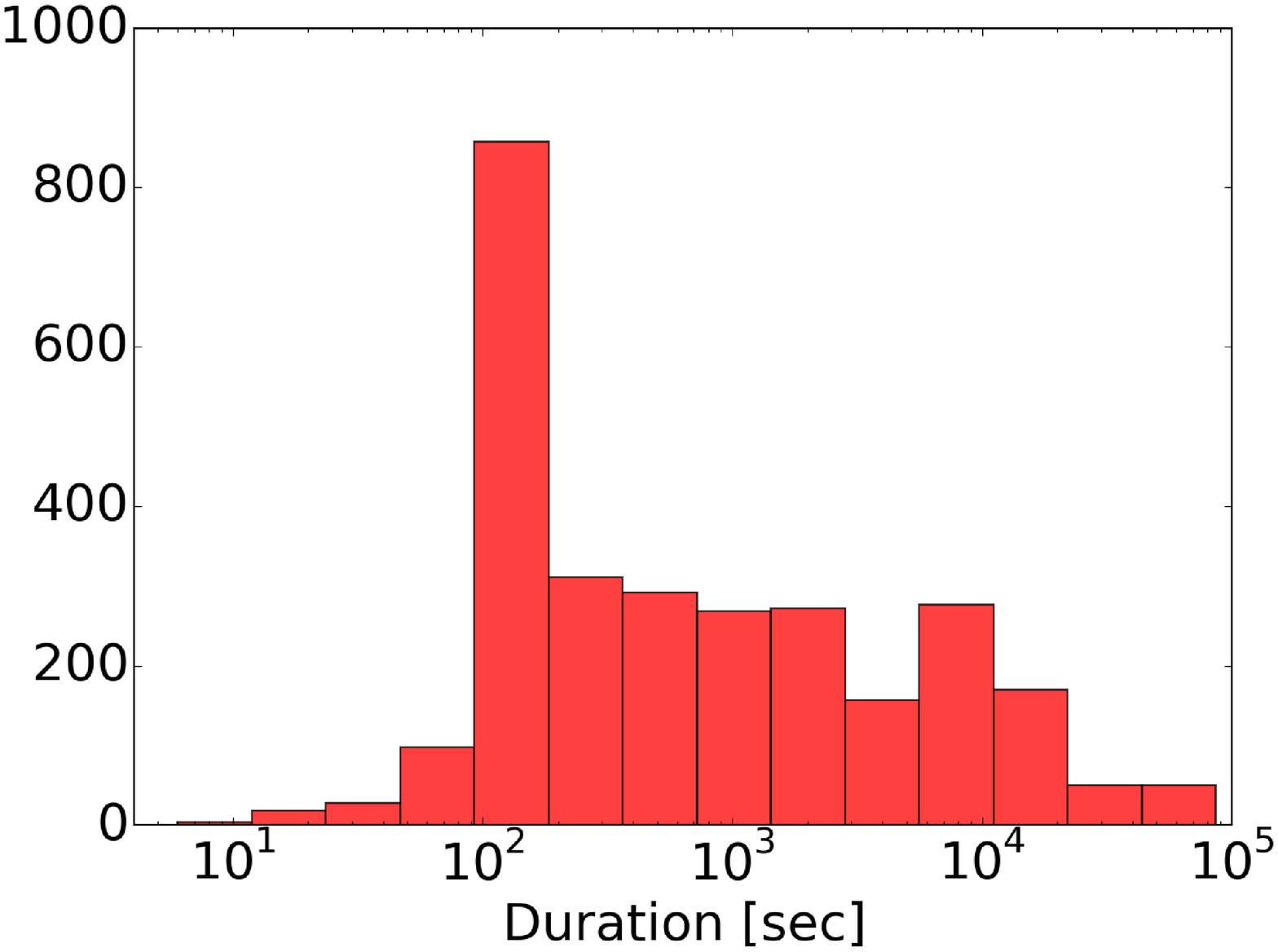}{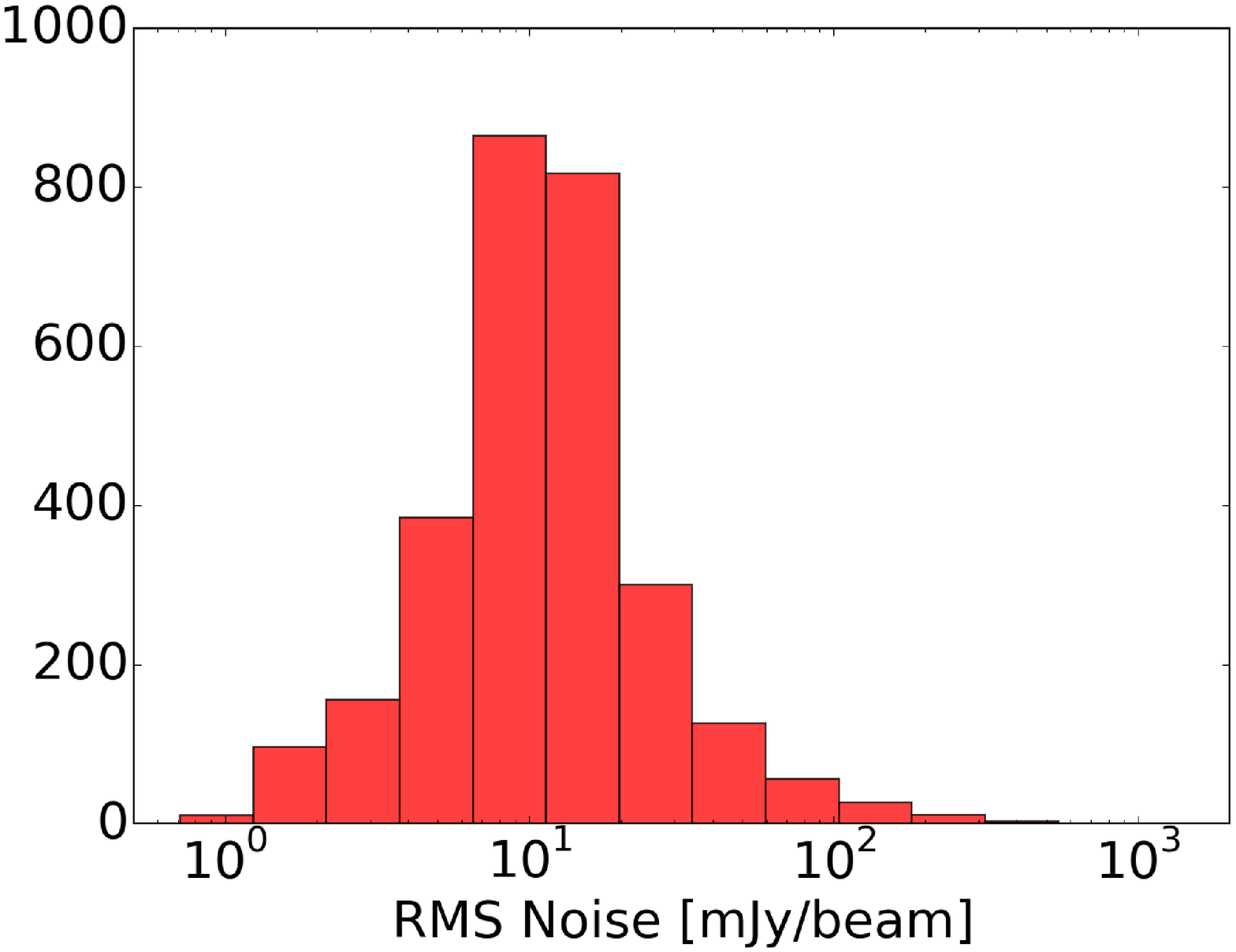}
\caption{The distribution of duration times (left) and the rms noise (right) in images for the 2799 individual daily VLA pointings.}
\label{fig2}
\end{figure}
\clearpage

\begin{figure}
\epsscale{.85}
\plotone{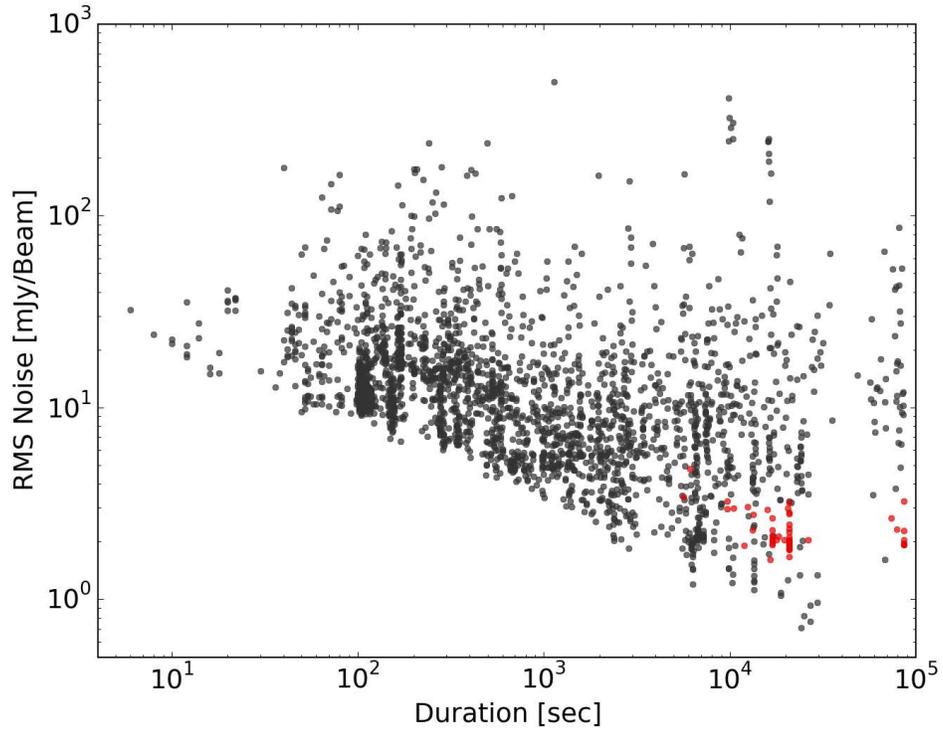}
\caption{RMS noise in the final pipeline processed images (see \S\ref{system}) as a function of the duration time. Plotted in black are the 2799 daily images; in red are the 55 images of the COSMOS field.}
\label{fig3}
\end{figure}
\clearpage

\begin{figure}
\plotone{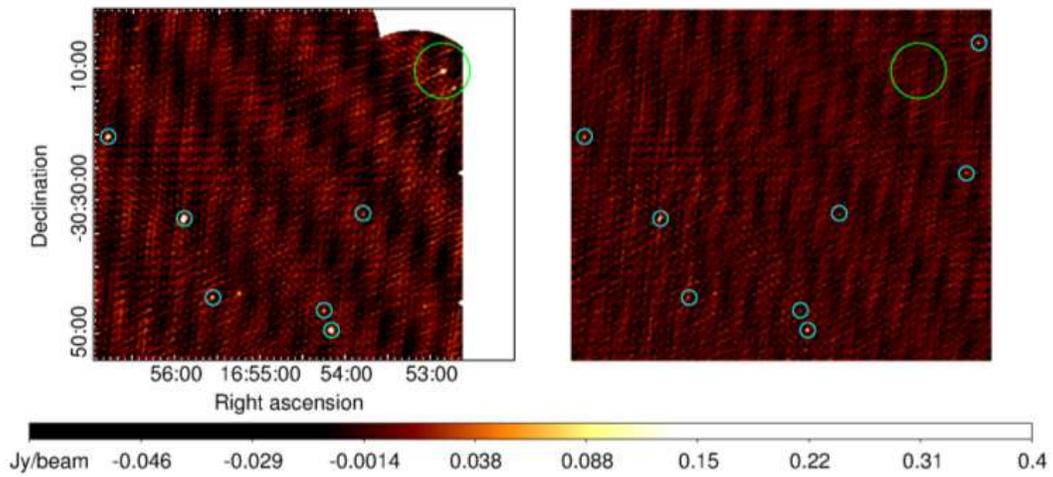}
\caption{Candidate transient source seen at the edge of a pipeline image at low resolution (left), and higher-resolution remake of the image which shows that it was an artifact (right). The angular resolution is $28.5\arcsec$ (left) and $12.3\arcsec$ (right), and the local rms noise in both images is 11 mJy/bm.  The candidate transient source is marked with a large green circle.  Small cyan circles indicate sources with NVSS counterparts.}
\label{Candidate1}
\end{figure}
\clearpage

\begin{figure}
\plotone{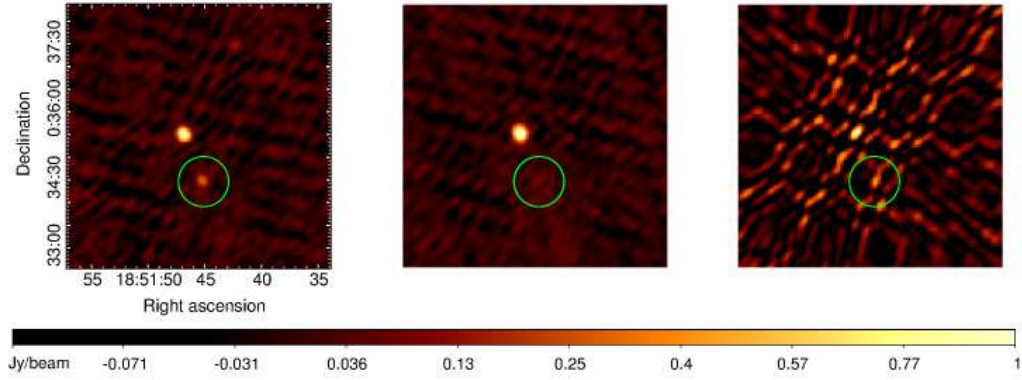}
\caption{Candidate $14\sigma$ transient source seen near a bright source in a pipeline image (left), and a hand-made image which excluded that position from the CLEAN area and shows that it was an artifact (center).  The local rms noise and angular resolution are 21 mJy/bm and $12.3\arcsec$ in both images.  The panel on the right shows the synthesized beam.  The candidate transient source position is circled in green on all three images.}
\label{Candidate2}
\end{figure}
\clearpage

\begin{figure}
\includegraphics[scale=0.70,angle=270]{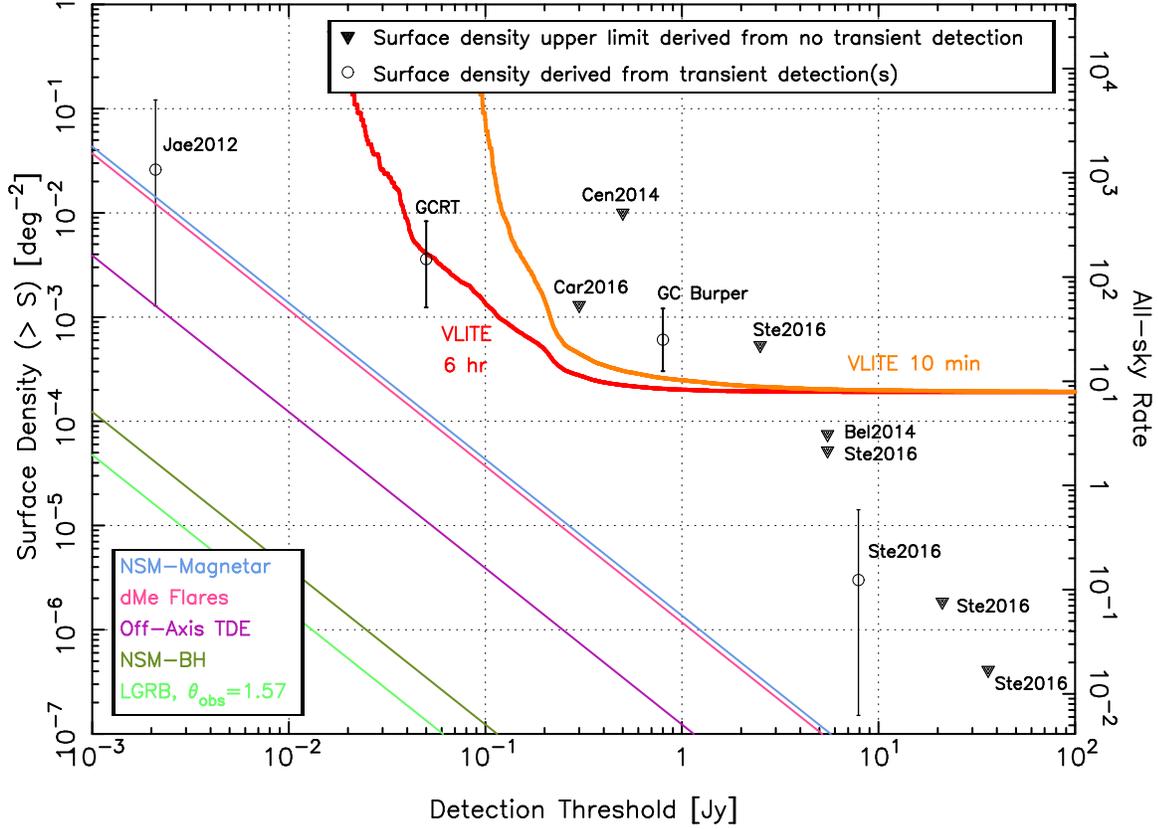}
\caption{Radio transient surface densities and all-sky rates are plotted against the survey detection threshold for blind transient searches at sub-GHz frequencies. Black triangles indicate upper limits from no transient detections. Round points with $95\%$ confidence bars are bounds from surveys with transient detections. Red and orange lines are the VLITE upper limits for 6 hour and 10 min timescale transients, respectively. Diagonal solid lines are estimated rates of the more common known transient source classes. The ``GC Burper'' surface density is based on the six Jy-level bursts detected from GCRT J1745$-$3009. The ``GCRT'' surface density plotted at 50 mJy by \citet{mhb+16} is based on all four GC transients detected, not taking into account the number of repeated events. The surveys included in this plot are: \citet{bmk+14} (Bel2014); \citet{cvw+14} (Car2016); \citet{cen14} (Cen2014); \citet{2012AJ....143...96J} (Jae2012); \citet{sfb+16} (Ste2016).}
\label{fig6}
\end{figure}
\clearpage

\begin{figure}
\includegraphics[scale=0.70,angle=270]{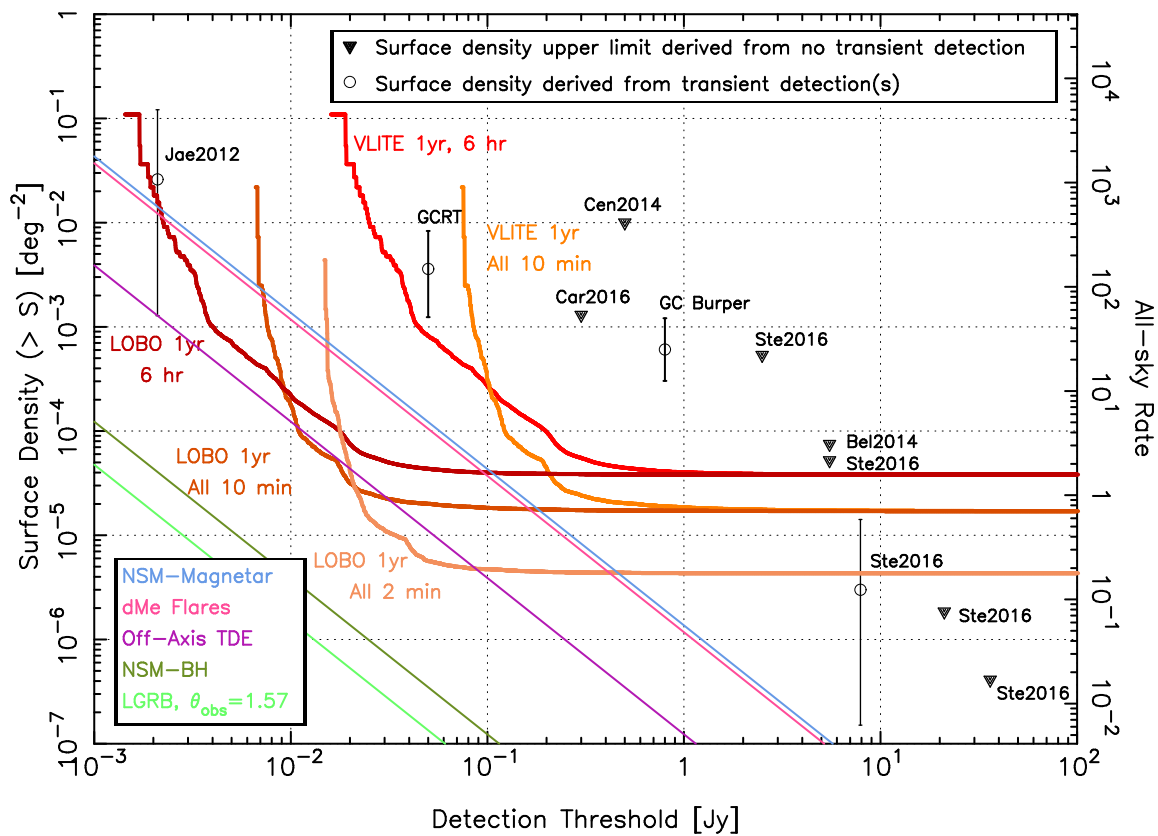}
\caption{Radio transient surface density and all-sky rate estimates for 1 year of VLITE data. The 10 min VLITE upper limit includes splitting long integration images into shorter timescales. Extending to lower detection thresholds are the 1 year projections for a full LOw Band Observatory (LOBO).}
\label{fig7}
\end{figure}
\clearpage


\clearpage

\begin{deluxetable}{ccccc}
\tablecaption{VLITE Limits for 10-min and 6-hr Transient Detection versus Peak Flux Density\label{tbl-1}}
\tablewidth{0pt}
\tablehead{ 
\colhead{Flux Density} & \colhead{Number of} & \colhead{$\Sigma_{10}$} & \colhead{Number of} & \colhead{$\Sigma_{6}$}\\
\colhead{(mJy)} & \colhead{10-min Epochs}\tablenotemark{a} &  \colhead{(deg$^{-2}$)} & \colhead{6-hr Epochs}\tablenotemark{a} & \colhead{(deg$^{-2}$)}\\
}
\startdata
25  & 0       & \omit                & 11  & $<5.0\times 10^{-2}$ \\ 
50  & 0       & \omit                & 133 & $<4.1\times 10^{-3}$ \\
100 & 7       & $<$0.1               & 404 & $<1.4\times 10^{-3}$ \\
150 & 132     & $<4.1\times 10^{-3}$ & 765 & $<7.1\times 10^{-4}$ \\
175 & 218     & $<2.5\times 10^{-3}$ & 917 & $<5.9\times 10^{-4}$ \\
200 & 376     & $<1.5\times 10^{-3}$ & 1114& $<4.9\times 10^{-4}$ \\
250 & 978     & $<5.6\times 10^{-4}$ & 1737& $<3.1\times 10^{-4}$ \\
300 & 1222    & $<4.5\times 10^{-4}$ & 1975& $<2.8\times 10^{-4}$ \\
400 & 1580    & $<3.5\times 10^{-4}$ & 2306& $<2.4\times 10^{-4}$ \\
600 & 1912    & $<2.9\times 10^{-4}$ & 2545& $<2.1\times 10^{-4}$ \\
1000 & 2199   & $<2.5\times 10^{-4}$ & 2711& $<2.0\times 10^{-4}$ \\
2000 & 2478   & $<2.2\times 10^{-4}$ & 2800& $<1.9\times 10^{-4}$ \\
10,000 & 2745 & $<2.4\times 10^{-4}$ & 2852& $<1.9\times 10^{-4}$ \\
\enddata
\tablenotetext{a}{Includes all epochs able to detect a transient of this timescale}
\end{deluxetable}
\end{document}